\documentclass[12pt,preprint]{aastex}

\begin{document}

\title{Estimating Power Spectrum of Sunyaev-Zeldovich Effect
from the Cross-Correlation between WMAP and 2MASS}

\author{Liang Cao, Jiren Liu and Li-Zhi Fang}

\affil{Department of Physics, University of Arizona,
Tucson, AZ 85721}

\begin{abstract}
We estimate the power spectrum of
SZ(Sunyaev-Zel'dovich)-effect-induced temperature fluctuations on
sub-degree scales by using the cross correlation between the
three-year WMAP maps and 2MASS galaxy distribution. We produced
the SZ effect maps by hydrodynamic simulation samples of the
$\Lambda$CDM model, and show that the SZ effect temperature
fluctuations are highly non-Gaussian. The PDF of the temperature
fluctuations has a long tail. More than $70\%$ power of the SZ
effect temperature fluctuations attributes to top $\sim 1\%$
wavelet modes (long tail events). On the other hand, the CMB
temperature fluctuations basically are Gaussian. Although the mean
power of CMB temperature fluctuations on sub-degree scales is much
higher than that of SZ effect map, the SZ effect temperature
fluctuations associated with top 2MASS clusters is comparable to
the power of CMB temperature fluctuations on the same scales.
Thus, from noisy WMAP maps, one can have a proper estimation of
the SZ effect power at the positions of the top 2MASS clusters.
The power spectrum given by these top wavelet modes is useful to
constrain the parameter of density fluctuations amplitude
$\sigma_8$. We find that the power spectrum of these top wavelet
modes of SZ effect on sub-degree scales basically is consistent
with the simulation maps produced with $\sigma_8=0.84$. The
simulation samples of $\sigma_8=0.74$ show, however, significant
deviation from detected SZ power spectrum. It can be ruled out
with confidence level 99\% if all other cosmological parameters
are the same as that given by the three-year WMAP results.

\end{abstract}


\section{Introduction}

The thermal Sunyaev-Zel'dovich (SZ) effect is due to the inverse
Compton scattering of the cosmic microwave background (CMB)
photons by hot electrons. This effect is proportional to the
pressure of the electron gases, $p =k_B nT$, where $n$ and $T$
being the number density and temperature of electrons. Therefore,
SZ effect can be used as a measurement of the hot baryon gases of
the universe (e.g. Birkinshaw 1999; Carlstrom et al. 2002). Since
clusters are the hosts of high pressure gases, the angular power
spectrum of the SZ effect is very sensitive to the amplitude of
mass density fluctuations parameter $\sigma_8$ (Refregier et al.
2000; Bond et al. 2002). The amplitude of the SZ effect angular
power spectrum is found to be proportional to
$\sigma_8^{7}(\Omega_bh)^2$ and is almost independent of other
cosmological parameters (e.g. Seljak et al. 2001; Komatsu \&
Seljak 2002). Therefore, the amplitude of the SZ effect power
spectrum can be used as an effective constraint on the parameter
$\sigma_8$. The SZ effect power spectrum on scales of a few
arcminutes has been used for this issue (Mason et al.2003;
Readhead et al. 2004; Dawson et al. 2006).

Recently released WMAP three-year data (WMAP III) refines most
results of cosmological parameters given by the WMAP first year
data. They found $\sigma_8=0.74\pm 0.05$ (Spergel et al. 2006),
which is significantly lower than $\sigma_8=0.84\pm 0.04$ of the
first year data. The new number of $\sigma_8$ is a challenge to the
cosmological parameter determined with samples of galaxies and
galaxy clusters, most of which yield $\sigma_8 \simeq 0.9-1$ if the
matter content of the universe $\Omega_m \leq 0.3$ (Reiprich \&
B{\"o}hringer 2002; Hoekstra et al. 2002; Refregier, et. al. 2002;
Van Waerbeke et. al. 2002; Bacon et al. 2003; Bahcall \& Bode 2003;
Seljak, et al. 2005; Viel  \& Haehnelt  2006; McCarthy et al. 2006;
Cai et al. 2006). This problem motivates us to study the constraint
on $\sigma_8$ given by the power spectrum of the SZ effect.

We try to extract the information of the SZ effect power spectrum on
sub-degree scales from the WMAP III data themselves. At the first
glance, it seems to be hopeless to extract such information from
noisy WMAP III data, because the power of SZ effect induced
temperature fluctuations on sub-degree scales is much less than that
of CMB temperature fluctuations (e.g. Cooray et al. 2004). A direct
analysis on the WMAP data also shows that the SZ effect contribution
to the CMB fluctuations on scale of the first acoustic peak should
be less than 2\% (Huffenberger et al. 2004; Spergel et al. 2006).

This problem can be solved if considering that the SZ effect map is
highly non-Gaussian, while CMB is Gaussian. Although the mean power
of the thermal SZ effect on sub-degree scales is no more than $2\%$
of the power of CMB, the SZ temperature fluctuations at local
modes associated with clusters are comparable to the CMB temperature
fluctuations. Thus, the SZ temperature fluctuations power on sub-degree
scales would be detectable from noisy WMAP maps if one can identify such
local modes. To implement this detection, we developed a DWT(discrete
wavelet transform) algorithm for the power spectrum estimation.
The DWT analysis is powerful to pick up weak but non-Gaussian signals
from Gaussian background (Donoho 1995).

We will use the cross-correlation between the WMAP III data and
2MASS galaxies. Since the redshift of 2MASS XSC galaxies is up to
$z\simeq 0.1$, at which clusters are on angular scales of
sub-degree. Moreover, the thermal SZ effect on angular scales of
clusters has been significantly identified with the 2nd and 4th
order cross-correlations between the first year WMAP data and 2MASS
XSC galaxies(Afshordi et al. 2004, 2005; Myers et al. 2004; Cao et
al. 2006). The WMAP-2MASS cross-correlation would be effective to
extract the SZ power on sub-degree scales from the WMAP III maps.

The paper is organized as follow. In \S 2, we analyze the
non-Gaussian features of the SZ effect maps generated with
cosmological hydrodynamic simulation. \S 3 describes the
identification of the SZ effect of 2MASS clusters from the
WMAP-2MASS cross correlation with the wavelet modes. \S 4 first
describes the algorithm of estimating the SZ power spectrum from the
WMAP III maps, and then presents the result of the SZ effect
constraint on $\sigma_8$. The discussion and conclusion are given in
\S 5.

\section{The power spectrum of SZ effect fluctuations}

\subsection{Maps of SZ effect}

To develop the method of estimating power spectrum, we first study
the statistical properties of the SZ effect temperature
fluctuations. We produce the SZ effect maps with the WIGEON (Weno
for Intergalactic medium and Galaxy Evolution and formatiON) code,
which is a cosmological hydrodynamic $N$-body code based on the
Weighted Essentially Non-Oscillatory (WENO) algorithm (Harten et
al. 1986; Liu et al. 1994; Jiang \& Shu 1996; Shu 1998; Fedkiw,
Sapiro \& Shu 2003; Shu 2003). The details of this code can be
found at Feng et al. (2004). Some WIGEON samples have already been
used in cosmological studies (e.g. He et al. 2004, 2006; Zhang et
al. 2006; Liu et al. 2006).

The samples are simulated in a 100 $h^{-1}$ Mpc cubic box with
512$^3$ meshes for gas, and equal number of particles for dark
matter. We use the standard $\Lambda$CDM model, which is specified
by the matter density parameter $\Omega_{\rm m}=0.27$, baryon
density parameter $\Omega_{\rm b}=0.044$, cosmological constant
$\Omega_{\Lambda}=0.73$, Hubble constant $h=0.71$. In order to
test the effect of the amplitude of density fluctuations, we
produce two samples with $\sigma_8=0.74$ and 0.84, respectively.
The ratio of specific heats is $\gamma=5/3$. The transfer function
is calculated using CMBFAST (Seljak \& Zaldarriaga 1996). In order
to correct the underestimate of rare massive galaxy clusters due
to the finite box size, the initial conditions are generated by
sampling a convolved power spectrum following the method of Pen
(1997) and Sirko (2006), which accurately model the real-space
statistics, such as $\sigma_8$.

The atomic processes including ionization, cooling and heating are
modeled as the method in Theuns et al. (1998). We take a
primordial composition of H and He ($X$=0.76, $Y$=0.24) and use an
ionizing background model of Haardt \& Madau (1996). Star
formation and their feedback due to supernova explosions and AGN
activities were not taken into account.

Since the redshift of 2MASS galaxies is up to 0.1, we produce the
SZ maps by integrating the pressure of gas from $z=0$ to 0.1. The
simulation data is stored for every redshifts corresponding to
comoving distance $n\times$100$h^{-1}$Mpc, $n$ being integral.
Three simulation boxes are stacked together and integrated along
different angular direction to compose a SZ map. Two 2-D maps of
the Compton parameter $y$ for $\sigma_8=0.74$ and 0.84 are shown
in Figure 1. Both maps have a size of $19.1^{\circ}\times
19.1^{\circ}$, with resolution of 512$\times$512. We produce 40
independent maps. The total area, $40\times (19.1^{\circ})^2$, is
comparable with the area of 2MASS data we used (\S 3.1).

Figure 2 gives the angular power spectrum for two samples with
$\sigma_8=0.74$ and 0.84, respectively. We see that the amplitude
of the power spectrum is significantly dependent on $\sigma_8$.
Comparing Figure 2 with other simulations (e.g. Refregier et al.
2000; Seljak et al. 2001; Komatsu \& Seljak 2002), our result
shows about the same power as others on scales equal to and larger
than $0^{\circ}.5$, while it is much lower than others on scales
$\ll 0^{\circ}.5$. This is expected, as the power on scales less
than $0^{\circ}.5$ mainly comes from clusters with redshift $>
0.1$. If considering all the contribution from $z=6$, our
simulation samples show the same behavior as others, i.e.,
declining around $0^{\circ}.1$.

\subsection{The DWT power spectrum of SZ effect map}

As shown by Figure 1, the value of $y$ is very small in most
space, while high value of $y$ appears only in small and localized
areas. One can then take the advantage of DWT analysis to describe
the field $y({\bf x})$ (${\bf x}=(x_1,x_2)$) with DWT
decomposition. The DWT variables of a map $y(x_1,x_2)$ in the area
of $0\leq x_1\leq L_1$ and $0\leq x_2\leq L_2$ are given by the
decompositions of scaling functions $\phi_{\bf j,l}({\bf x})$ and
wavelet functions $\psi_{\bf j,l}({\bf x})$. We use Daubechies 4
wavelet (Daubechies (1992), Fang \& Thews (1998). The functions
$\phi_{\bf j,l}({\bf x})$ and $\psi_{\bf j,l}({\bf x})$ can also
be found in Yang et al. (2001).

Briefly, The DWT scaling function, $\phi_{\bf j,l}({\bf x})$, is a
window function for cells (modes) in the area $L_1l_1/2^{j_1} < x_1<
L_1(l_1+1)/2^{j_1}$ and $L_2l_2/2^{j_2} < x_2< L_2(l_2+1)/2^{j_2}$.
The scale index ${\bf j}=(j_1,j_2)$, $j_1$ and $j_2$ can be any
positive integer. The position index ${\bf l}=(l_1,l_2)$, $l_1$ and
$l_2$ can be 0, 1... $2^{j_1}-1$, and 0, 1... $2^{j_2}-1$,
respectively. On the other hand, the wavelet function is admissible,
satisfying $\int \psi_{\bf j,l}({\bf x})d{\bf x}=0$.

For the field $y({\bf x})$, the DWT variables are
\begin{equation}
y_{\bf j,l}=\frac{1 }
 {\int\phi_{\bf j,l}({\bf x})d{\bf x}}
  \int y({\bf x})\phi_{\bf j,l}({\bf x})d{\bf x},
\end{equation}
\begin{equation}
\tilde{\epsilon}^y_{\bf j,l}= \int y({\bf x}) \psi_{\bf j,l}({\bf
x})d{\bf x},
\end{equation}
Since $\phi_{\bf j,l}({\bf x})$ is a window function of cell
$\bf(j,l)$, $y_{\bf j,l}$ is the mean of $y(\bf x)$ in cell
$\bf(j,l)$. Because $\psi_{\bf j,l}({\bf x})$ is admissible,
$\tilde{\epsilon}^y_{\bf j,l}$ measures the fluctuation of field
$y({\bf x})$ with respect to the mean, $y_{\bf j,l}$, in the
mode $({\bf j,l})$. For a given ${\bf j}$, the mean of
$\langle \tilde{\epsilon}^y_{\bf j,l} \rangle$ over all modes is
zero (Fang \& Thews 1998). Since the basis $\psi_{\bf j,l}$ are
orthogonal and complete, the DWT variables of the field $y({\bf x})$
do not lost information and cause false correlations.

With the DWT variables, the power spectrum of the 2-D field
$y({\bf x})$ can be measured as (Pando \& Fang 1998; Fang \& Feng
2000):
\begin{equation}
P_{\bf j}=\langle |\tilde{\epsilon}_{\bf j,l}|^2\rangle\equiv
   \frac{1}{2^{j_1+j_2}}
\sum_{l_1=0;l_2=0}^{2^{j_1}-1; 2^{j_2}-1}
     |\tilde{\epsilon}^y_{j_1,j_2;l_1,l_2}|^2,
\end{equation}
$|\tilde{\epsilon}^y_{j_1,j_2;l_1,l_2}|^2$ is the power at mode
$({\bf j,l})=(j_1,j_2;l_1,l_2)$. Therefore, the DWT power spectrum
$P_{\bf j}$ actually is the mean of local powers
$|\tilde{\epsilon}_{\bf j,l}|^2$ over all $2^{j_1+j_2}$ modes on
scale ${\bf j}$. It has been shown that the DWT power spectrum can
be expressed by the Fourier power spectrum as (Fang \& Feng 2000;
He et al. 2005a):
\begin{equation}
P_{\bf j} =  \frac{1}{2^{j_1+j_2}}
  \sum_{n_1 = - \infty}^{\infty}
  \sum_{n_2 = - \infty}^{\infty}
 |\hat{\psi}(n_1/2^{j_1})\hat{\psi}(n_2/2^{j_2})|^2 P(n_1,n_2),
\end{equation}
where $\hat{\psi}(n)$ is the Fourier transform of the basic
wavelet $\psi(x)$. Because we have the normalized relation
$\sum_{n_1,n_2}|\hat{\psi}(n_1/2^{j_1})\hat{\psi}(n_2/2^{j_2})
|^2/2^{j_1+j_2}=1$, the DWT power spectrum $P_{\bf j}$ is equal to
the Fourier power spectrum  banded on scales around
$(L_1L_2/2^{j_1+j_2})^{1/2}$. Figure 3 shows both the angular
power spectrum $C_l$ and DWT power spectrum $P_{\bf j}$ for the
simulation samples. The angular scale of $j$ is $19.1/2^j$ degree,
and therefore, $l\simeq 9.4\times 2^j$. Figure 3 shows that DWT
power spectrum is the same as the angular power spectrum $C_l$. We
will use the DWT analysis on angular scales of sub-degree, on
which the effect of spherical surface is not significant.

\subsection{Non-Gaussianity of SZ effect map}

The DWT power spectrum can effectively detect the non-Gaussian
feature of a field (Jamkhedkar et al. 2003). Taking this advantage,
one can reveal the non-Gaussianity of the SZ effect field $y({\bf
x})$ with the DWT power spectrum of the SZ field $y({\bf x})$
defined as
\begin{equation}
P^{\rm{top}}_{\bf j}=\frac{1}{2^{j_1+j_2}} \sum_{\rm{top}}
|\tilde{\epsilon}_{j_1,j_2;l_1,l_2}|^2,
\end{equation}
where summation of ``top'' goes over only modes having high values
$|\tilde{\epsilon}_{j_1,j_2;l_1,l_2}|^2$ among the $2^{j_1+j_2}$
modes for a given ${\bf j}$. In Figure 4, we plot the $P_{\bf j}$
and $P^{\rm{top}}_{\bf j}$ of the simulation samples, in which top
power spectrum, $P^{\rm{top}}_{\bf j}$, is given by top 1\%,
0.3\%, 0.1\% among the $2^{j_1+j_2}$ modes. We see that all the
power spectra $P^{\rm{top}}_{\bf j}$ are close to $P_{\bf j}$. For
samples with $\sigma_8=0.84$, there is more than $70\%$, $50\%$,
$35\%$ power attributes from top 1\%, 0.3\%, 0.1\% modes,
respectively. That is, the power $
|\tilde{\epsilon}^{\rm{top}}_{\bf j,l}|^2$ of each mode among the
top 1\% modes should be as high as about 70 times of the mean
$\langle{|\tilde{\epsilon}_{\bf j,l}|^2}\rangle=P_{\bf j}$. These
results show that the field $y({\bf x})$ is highly non-Gaussian.
From Figure 4, one can see that for sample with $\sigma_8=0.74$,
all the power spectra $P^{\rm{top}}_{\bf j}$ of top 1\%, 0.3\%,
0.1\% are less close to $P_{\bf j}$. This is because the
clustering of $\sigma_8=0.84$ sample is stronger than that of
$\sigma_8=0.74$.

Because $\langle \tilde{\epsilon}_{\bf j,l} \rangle =0$, the value
of $\langle{|\tilde{\epsilon}_{\bf j,l}|^2}\rangle^{1/2}$ actually
is the standard deviation($\sigma$) of the probability
distribution function (PDF) of $\tilde{\epsilon}_{\bf j,l}$. Thus,
the events of 70$\langle{|\tilde{\epsilon}_{\bf j,l}|^2}\rangle$
actually are $\sim$ 8$\sigma$ events. All the top 1\% wavelet
modes are 8$\sigma$ events. All the top 0.3\% and 0.1\% modes
correspond, respectively, to $\sim
170$$\langle{|\tilde{\epsilon}_{\bf j,l}|^2}\rangle$, and $\sim
350$$\langle{|\tilde{\epsilon}_{\bf j,l}|^2}\rangle$, or 13
$\sigma$ and 19 $\sigma$ events. For a Gaussian field, the 10
$\sigma$ events should be no more than $10^{-23}$. Therefore, the
PDF of $\tilde{\epsilon}_{\bf j,l}$ has a very long tail. In other
words, a significant part of the SZ effect DWT power spectrum is
given by the long tail events. These long tail events make the SZ
effect on sub-degree scales to be detectable. Although the mean of
SZ power $\tilde{\epsilon}_{\bf j,l}$ is less than about 2\% of
the power of CMB, for long tail events, the SZ power is comparable
with that of CMB.

One can estimate the power spectrum of SZ effect temperature
fluctuations with the top modes (long tail event) of
$\tilde{\epsilon}_{\bf j,l}$. For observed maps, however, not all
the top modes of $\tilde{\epsilon}_{\bf j,l}$ are given by SZ
effect. To reduce this uncertainty, we add a condition that the
useful modes $({\bf j,l})$ for top $\tilde{\epsilon}_{\bf j,l}$
should also be the modes significantly showing SZ effect by the
WMAP-2MASS cross correlation. It is equal to assume that the power
of SZ effect temperature fluctuations mainly comes from clusters
(Komatsu \& Kitayama 1999; Komatsu \& Seljak 2002). We check this
assumption with the relation between variables $y_{\bf j,l}$ and
$\tilde{\epsilon}_{\bf j,l}$. We found that the top modes of
$|\tilde{\epsilon}_{\bf j,l}|$ are largely coincident with the top
modes of the Compton parameter $y_{\bf j,l}$. For sample of
$\sigma_8=0.84$, $84\%$ top modes of $y_{\bf j,l}$ on the angular
scale of $0^{\circ}.5$ are coincident with the top modes of
$|\tilde{\epsilon}_{\bf j,l}|$ on the same scale. This result
indicates that modes with high $y_{\bf j,l}$ generally are also the
modes of high local power of SZ effect temperature fluctuations.
High $y_{\bf j,l}$ generally is related to clusters.

The fact that the coincidence between top $y_{\bf j,l}$ and top
$\tilde{\epsilon}_{\bf j,l}$ is not 100\% is due partially to the
so-called warm-hot intergalactic medium (WHIM) (Dave et al 2001; He
et al. 2004, 2005b). It shows that a significant fraction of hot
baryon gases ($T>10^5$) are not located in collapsed structures with
high density contrast $\rho/\bar{\rho}> 100$, but in the areas with
a median overdensity $10< \rho/\bar{\rho} < 30$. The WHIM is also
the source of SZ effect. Nevertheless, the major part of top modes
of $|\tilde{\epsilon}_{\bf j,l}|$ is associated with clusters.
Therefore, a proper method to estimate the SZ power spectrum would
be to identify the modes with top numbers of both $y_{\bf j,l}$ and
$|\tilde{\epsilon}_{\bf j,l}|$. The details of this method will be
given in the following two sections.

\section{SZ effect identified from WMAP-2MASS correlation}

\subsection{Data}

The data of galaxies in the 2MASS extended source catalog (XSC,
Jarrett et al. 2000) used for cross-correlation analysis is the
same as Guo et al. (2004) and Cao et al. (2006). That is, we use
galaxies of ${\rm K\_m\_k20fe}$, which measures the magnitude
inside a elliptical isophote with surface brightness of 20 mag
${\rm arcsec^{-2}}$ in $K_s$-band. There are approximately 1.6
million extended objects with $K_s<14.3$. To avoid the contaminant
of stars, we use a latitude cut of $|b| > 10^{\circ}$. We also
removed a small number of bright ($K_s<9$) sources  by the
parameters of the XSC confusion flag (${\rm cc\_flag}$) and visual
verification score for source (${\rm vc}$). To eliminate duplicate
sources and have a uniform sample, we use ${\rm use\_src=1}$ and
${\rm dup\_src=0}$ \footnote{The notations of the 2MASS parameters
used in this paragraph are from the list shown in the 2MASS Web
site http://www.ipac.caltech.edu/2mass/releases/allsky/doc.}. We
also use a cut of $10.0<K_s<14.0$ to ensure the sample is complete
greater than $90\%$. This sample contains 987,125 galaxies with
redshift up to $z \sim 0.1$.

For the CMB temperature fluctuations, we use the full resolution
coadded 3 year sky map (Hinshaw et al. 2006) for $W$ band. The
maps have had the ILC(Internal Linear Combination) estimate of the
CMB signal to highlight the foreground emissions(Hinshaw et al.
2006). We also use the three annual maps to construct three
difference maps between the 1st and 2nd, 2nd and 3rd, 3rd and 1st
year, which are useful to check statistical significance.

As in Cao et al. (2006), we subject both the 2MASS and WMAP maps
to an equal-area projection. We then have 2-D maps of galaxy
number density, $n({\bf x})$, and CMB temperature fluctuations,
$\Delta T({\bf x})$ with coordinate ${\bf x}=(x_1,x_2)$. As has
been shown, the equal-area projection will not affect statistical
analysis on sub-degree scales, but helpful for using the DWT
method. We analyze the cross correlation between $n({\bf x})$ and
$\Delta T({\bf x})$ in two $123^{\circ}.88 \times 123^{\circ}.88$
areas, which are, respectively, in the northern and southern sky.
The 2MASS XSC galaxies are resolved to 10 arcsec. The beam size of
WMAP $W$ band map is
$0^{\circ}.22$\footnote{http://lambda.gsfc.nasa.gov/product},
while the scale of galaxy clusters at the median redshift of the
2MASS samples is $0^{\circ}.5$.

In order to apply the method of DWT power spectrum (\S 2.2), we also
use the DWT variables to describe the maps $n({\bf x})$ and
$\Delta T({\bf x})$. They are wavelet variables
$\tilde{\epsilon}^g_{\bf j,l}=
\int n({\bf x}) \psi_{\bf j,l}({\bf x})d{\bf x}$,
$\tilde{\epsilon}^T_{\bf j,l} = \int \Delta T({\bf x})\psi_{\bf
j,l}({\bf x})d{\bf x}$, and scaling function variables
$n_{\bf j,l}=\int n({\bf x})\phi_{\bf j,l}({\bf x})d{\bf x}/
\int\phi_{\bf j,l}({\bf x})d{\bf x}$,
$\Delta T_{\bf j,l}=
 \int \Delta T({\bf x})\phi_{\bf j,l}({\bf x})d{\bf x}/
\int\phi_{\bf j,l}({\bf x})d{\bf x}$. The mode $({\bf j,l})$
occupy a area of $123^{\circ}.88/2^{j_1} \times
123^{\circ}.88/2^{j_2}$ at position around
$[l_1(123^{\circ}.88)/2^{j_1}, l_2123^{\circ}.88)/2^{j_1}]$. The
angular distance between modes ${\bf l}$ and ${\bf l'}$ at scale
$j$ is given by $\theta =123.88^{\circ}|{\bf l-l'}|/2^j$.

The variables $\tilde{\epsilon}^T_{\bf j,l}$ and
$\tilde{\epsilon}^g_{\bf j,l}$ describe, respectively, the
fluctuations of the CMB temperature and galaxy number density in
the mode $({\bf j, l})$, while the variables $\Delta T_{\bf j,l}$
and $n_{\bf j,l}$ describe, respectively, the mean temperature and
the mean number density of galaxies in the mode $({\bf j,l})$.
Therefore, $n_{\bf j,l}$ and $\Delta T_{\bf j,l}$ can be seen as
the maps of galaxies and CMB temperature functions smoothed on
scale ${\bf j}$.

\subsection{DWT clusters}

We picked up top clusters on scale ${\bf j}$ by the top members of
$n_{\bf j,l}$. These clusters are called DWT clusters. The details
of $n_{\bf j,l}$ identification of clusters have been studied with
simulation and real samples (Xu et al. 1999, 2000). It showed that
the clusters identified by DWT $n_{\bf j,l}$ are statistically the
same as the clusters identified by the friend-of-friend method if
the mean size of the friend-of-friend identified clusters is the
same as DWT clusters.  The clusters identified by the
friend-of-friend method usually have very irregular shapes (Jing \&
Fang 1994) and are inconvenient to estimate the statistical
significance of the cross correlation. On the other hand, the DWT
variables of 2MASS galaxies $n_{\bf j,l}$ and WMAP map $\Delta
T_{j,l}$ are in the same cell, the statistical significance of the
$n_{\bf j,l}$-$\Delta T_{\bf j,l}$ cross correlation is unambiguous.
Scaling functions are orthogonal with each other, different DWT
clusters consist of different galaxies. This is also necessary for
statistical analysis which we will take below. With this method, we
identify top clusters on scales ${\bf j}=(8,8)$, corresponding to
scale $\simeq 1.8$ h$^{-1}$ Mpc of the 2MASS samples.

\subsection{Thermal SZ effect of 2MASS clusters}

The SZ effect of 2MASS galaxies can be measured with the
cross-correlation between the WMAP and 2MASS clusters defined as
Cao et al.(2006):
\begin{equation}
\Delta T^{\rm wmap-2mass}(|{\bf l-l'}|)=
   \langle C_{\bf j,l}\Delta T_{\bf j',l'}\rangle,
\end{equation}
where the variable $C_{\bf j,l}$ is given by:
\begin{equation}
C_{\bf j,l}= \left \{ \begin{array}{ll}
                    1 & {\rm if \ ({\bf j,l}) \ is \ a \ cell \ of
                                \ identified \ clusters}, \\
                    0 & {\rm otherwise}.
                  \end{array}  \right .
\end{equation}
The average $\langle ...\rangle$ covers all the DWT clusters on
scale ${\bf j}$. Therefore, $\Delta T^{\rm wmap-2mass}(|{\bf
l-l'}|)$ measures the mean of the CMB temperature fluctuations on
a distance $|{\bf l-l'}|$ from identified DWT clusters of 2MASS
data. When $|{\bf l-l'}|\gg 1$, $\Delta T^{\rm wmap-2mass}(|{\bf
l-l'}|)$ actually is the mean of $\Delta T$ on a large area, and
therefore, it should approach to zero. Figure 5 presents the cross
correlation $\Delta T^{\rm wmap-2mass}(|{\bf l-l'}|)$ between top
100 ${\bf j}=(8,8)$ DWT clusters and WMAP map, in which the scales
${\bf j'}$ of  $\Delta T_{\bf j',l'}$ are taken to be ${\bf
j'}=(8,8)$, ${\bf j'}=(8,7)$, and ${\bf j'}=(7,7)$ from top to
bottom. In all case $\Delta T_{\bf j',l'}$ is from the full
resolution coadded 3 year WMAP map of $W$ band. The error bars are
given by the 1-$\sigma$ standard deviation of the 100 CMB
simulations.

Figure 5 shows typical anti-correlation of SZ effect. At the
position of DWT clusters, i.e. $|\bf l-\bf l'|=0$, the mean of
temperature change $\Delta T_{\bf j',l'}$ is always negative,
which is expected from the thermal SZ effect on $W$ band. For
${\bf j'}=(8,8)$, $\Delta T^{\rm wmap-2mass}(|{\bf l-
l'}|=0)\simeq -25 \pm 15$ $\mu$K, while for ${\bf j'}=(8,7)$ and
${\bf j'}=(7,7)$, the temperature decreases are $-20 \pm 16$
$\mu$K and $-18 \pm 14$ $\mu$K respectively. These results are
consistent with that given by cross-correlation of the first year
WMAP map and 2MASS galaxies (Myers et al. 2004;
Hern\'andez-Monteagudo et al. 2004; Cao et al. 2006).

To further check the SZ signals, we calculate the cross-correlation
of eq.(6) by using the maps given by the difference between the 1st
and 2nd, 2nd and 3rd, 3rd and 1st year. The results are shown in
Figure 6. All the anti-correlations shown in Figures 5 completely
disappear in Figure 6. It confirms that the temperature decreases at
the positions of the DWT clusters in the three annual maps are about
the same and not due to noise.

To view the clusters richness dependence, Figure 7 plots $\Delta
T^{\rm wmap-2mass}(|{\bf l-l'}|=0)$ against the number of top
clusters, $n$. As expected, the higher the richness, the stronger
the SZ effect. $\Delta T^{\rm wmap-2mass}(|{\bf l-l'}|=0)$ rapidly
decreases with the number of top clusters, and approaches to the
noise level at $n>300$. We should, however, point out that the
positive correlation between the clusters richness and SZ effect
is statistical. That is, the order of clusters richness is not,
one-by-one, corresponding to the order of the amount of SZ effect
temperature decrease. Consequently, the relation of richness-SZ
effect in the range $n<100$ is not very stable. When $n \geq 100$,
the relation of richness-SZ effect, as shown in Figure 6 is very
stable. Therefore, in all the statistics below, the number of top
clusters should be $>100$.

In the DWT analysis, wavelet modes are orthogonal from each other.
Therefore, the area fraction of top 300 ${\bf j}=(8,8)$ DWT
clusters is $300/(2\times256^2)=0.23\%$. As shown in \S 2, the
mean power of SZ effect of top 0.3\% modes would be 170 times
higher than mean power of total mode. On the other hand, on
sub-degree scales, the CMB power is higher than the SZ effect
power by a factor $\sim 10^2$. Therefore, we can expect that the
mean of SZ temperature decreases of the top 0.3\% modes should be
larger than the CMB temperature fluctuations by a factor $\sim
\sqrt{170/100}\simeq 1.3$. Figure 8 plots the richness-dependence
of the S/N ratio, i.e. the ratio between SZ effect temperature
decrease and the $1-\sigma$ error bar shown in Figure 7. Since the
error bar of Figure 7 is given by the simulation of CMB
temperature fluctuations, and therefore, it actually is to measure
the the CMB temperature fluctuations. We see from figure 8 that
the S/N ratio is about 1.3 - 1.6. This result is consistent with
the estimation mentioned above.

\section{Estimation of DWT power spectrum of SZ effect}

\subsection{Method}

The observed CMB temperature fluctuations are given by:
\begin{equation}
\Delta T({\bf x}) = \Delta T^{\rm cmb}({\bf x})+ \Delta T^{\rm
sz}({\bf x})+ \Delta T^{\rm second}({\bf x}),
\end{equation}
where $\Delta T^{\rm cmb}({\bf n})$ is the primeval temperature
fluctuations, $\Delta T^{\rm sz}({\bf x})$ is the SZ effect and
$\Delta T^{\rm second}({\bf n})$ is due to secondary effects other
than the SZ effect, such as the ISW effect and radio point
sources. Since the random fields $\Delta T^{\rm cmb}({\bf x})$,
$\Delta T^{\rm sz}({\bf x})$ and $\Delta T^{\rm second}({\bf x})$
basically are uncorrelated from each other, the DWT power
spectrum, $P_{\bf j}$, of the observed $\Delta T({\bf x})$
consists of three terms as:
\begin{equation}
P_{\bf j} = P^{\rm cmb}_{\bf j}+ P^{\rm sz}_{\bf j}+P^{\rm
second}_{\bf j},
\end{equation}
where
\begin{equation}
P^{\rm cmb, sz, second}_{\bf j}=\frac{1}{2^{j_1+j_2}}
\sum_{l_1=0;l_2=0}^{2^{j_1}-1; 2^{j_2}-1}
     \left |\tilde{\epsilon}^{\rm cmb, sz,
     second}_{j_1,j_2;l_1,l_2}\right|^2.
\end{equation}

Since $P^{\rm cmb}_{\bf j}$ is about $10^2$ times larger than
$P^{\rm sz}_{\bf j}$ on angular scale of $\simeq 0^{\circ}.5$, one
cannot extract $P^{\rm sz}_{\bf j}$ from noisy map $\Delta T({\bf
x})$ if the field $\Delta T^{\rm sz}({\bf x})$ is Gaussian. As
shown in \S 2, the SZ effect DWT power spectrum is largely
given by the top modes. The local power of the top DWT modes can
be as large as $10^2$ times of their average. Therefore, the SZ
effect local power of the top modes is comparable to $P^{\rm
cmb}_{\bf j}$. On the other hand, the primeval temperature
fluctuations $\Delta T^{\rm cmb}({\bf x})$ are Gaussian, and the
PDF of local temperature fluctuations $\tilde{\epsilon}^T_{\bf
j,l}$ is Gaussian with zero mean and standard deviation $(P^{\rm
cmb}_{\bf j})^{1/2}$. The primeval temperature fluctuations map
does not have the modes with very high local power. Thus, the
local power of modes associated with top clusters would be
measurable from the noisy map of $\Delta T({\bf x})$.

The measurability of power of top modes can be further tested as
follows. Subjecting eq.(8) to a wavelet transform, we have:
\begin{equation}
\tilde{\epsilon}_{\bf j,l}= \tilde{\epsilon}^{\rm cmb}_{\bf j,l}+
\tilde{\epsilon}^{\rm sz}_{\bf j,l}+\tilde{\epsilon}^{\rm
second}_{\bf j,l},
\end{equation}
For most modes $({\bf j,l})$ on scales of clusters,
$|\tilde{\epsilon}^{\rm cmb}_{\bf j,l}|$ is much larger than
$|\tilde{\epsilon}^{\rm sz}_{\bf j,l}|$, and therefore, the latter
is negligible. But for modes corresponding to the top clusters,
$|\tilde{\epsilon}^{\rm sz}_{\bf j,l}|$ would be comparable to
$|\tilde{\epsilon}^{\rm cmb}_{\bf j,l}|$. On these modes, the term
$|\tilde{\epsilon}^{\rm second}_{\bf j,l}|$ is also negligible, as
statistically the positions of secondary effect are not coincident
to top clusters. Thus, on these modes, we have:
\begin{equation}
 |\tilde{\epsilon}_{\bf j,l}|^2=
|\tilde{\epsilon}^{\rm cmb}_{\bf j,l}|^2+ |\tilde{\epsilon}^{\rm
sz}_{\bf j,l}|^2,
\end{equation}
and therefore,
\begin{equation}
\frac{1}{N_{\rm top}}\sum_{\rm top}|\tilde{\epsilon}^{\rm sz}_{\bf
j,l}|^2 = \frac{1}{N_{\rm top}}\sum_{\rm
top}|\tilde{\epsilon}_{\bf j,l}|^2 -\frac{1}{N_{\rm top}}\sum_{\rm
top}|\tilde{\epsilon}^{\rm cmb}_{\bf j,l}|^2,
\end{equation}
where the summation ``top'' is over modes with top
$\tilde{\epsilon}_{\bf j,l}$ and also the top DWT clusters.
$N_{\rm top}$ is the number of top modes. Because CMB field is
statistically isotropic (uniform), Gaussian and uncorrelated with
2MASS data, the average of the primeval temperature fluctuations
power $|\tilde{\epsilon}^{\rm cmb}_{\bf j,l}|^2$ done by
$(1/N_{\rm top})\sum_{\rm top}$ should be about the same as the
average over all modes. We have then:
\begin{equation}
\sum_{\rm top}|\tilde{\epsilon}^{\rm cmb}_{\bf j,l}|^2\simeq
  N_{\rm top}\langle |\tilde{\epsilon}_{\bf j,l}|^2\rangle,
\end{equation}
where $\langle |\tilde{\epsilon}_{\bf j,l}|^2\rangle$ is the mean
of $|\tilde{\epsilon}_{\bf j,l}|^2$ over all modes on scale ${\bf
j}$. Thus, considering $|\tilde{\epsilon}^{\rm sz}_{\bf
j,l}|^2>0$, eq.(13) yields:
\begin{equation}
\frac{1}{N_{\rm top}}\sum_{\rm top}|\tilde{\epsilon}_{\bf j,l}|^2
> \langle |\tilde{\epsilon}_{\bf j,l}|^2\rangle,
\end{equation}
Eq.(15) requires that the mean local power over modes at top
clusters should be larger than the mean on all modes.

We test Eq.(15) by the cross correlation defined similar to
eq.(6):
\begin{equation}
\Upsilon_{\bf j'}(|{\bf l-l'}|)= \frac{
   \langle C_{\bf j,l}|\tilde{\epsilon}_{\bf j',l'}|^2 \rangle}
   {\langle |\tilde{\epsilon}_{\bf j',l'}|^2\rangle},
\end{equation}
where $C_{\bf j,l}$ is given by eq.(7). The numerator on the
r.h.s. of eq.(16) is to measure the mean local power of CMB
temperature fluctuations over the modes of DWT clusters on scale
${\bf j'}$. The denominator of eq.(16) is for normalization. The
inequality in eq.(15) requires $\Upsilon_{\bf j'}(|{\bf l-l'}|)>1$
at $|\bf l-\bf l'|=0$, i.e. the local power of the modes, which
shows significant SZ temperature decrease, should also show power
larger than the mean $\langle |\tilde{\epsilon}_{\bf
j',l'}|^2\rangle$ over the whole sky. The result of $\Upsilon_{\bf
j'}(|{\bf l-l'}|)$ is plotted in Figure 9, in which the scale
${\bf j'}$ of local power $|\tilde{\epsilon}_{\bf j',l'}|^2$ is
taken to be ${\bf j'}=(8,8)$, ${\bf j'}=(8,7)$ and ${\bf
j'}=(7,7)$. For each scale, we count 0.11\% modes with top
$|\tilde{\epsilon}_{\bf j',l'}|^2$, and all the counted modes are
located in the positions of the 300 top 2MASS clusters,
corresponding to 150, 75, and 38 top modes of the scales ${\bf
j'}=(8,8)$, ${\bf j'}=(8,7)$ and ${\bf j'}=(7,7)$. Figure 9 shows
that all the mean powers of CMB temperature fluctuations on the
three scales are significantly larger than the mean
$\langle|\tilde{\epsilon}_{\bf j',l'}|^2\rangle$. This power
excess is due to the SZ effect of the 2MASS clusters, which are
larger than the noise of WMAP III data.

\subsection{Results}

With these preparation, we can estimate the SZ effect power
spectrum given by modes associated with the top DWT clusters of
2MASS galaxies. It is,
\begin{equation}
P^{\rm top}_{\bf j}=\frac{1}{2^{j_1+j_2}} \sum_{\rm
top}|\tilde{\epsilon}^{\rm sz}_{\bf j,l}|^2,
\end{equation}
where the ``top'' is taken the same modes as shown in Figure 9.
Using eqs.(13) and (14), eq.(17) gives,
\begin{equation}
P^{\rm top}_{\bf j} \simeq \frac{1}{2^{j_1+j_2}}\left [ \sum_{\rm
top}
     |\tilde{\epsilon}_{\bf j,l}|^2-N_{\rm top}
\langle |\tilde{\epsilon}_{\bf j,l}|^2\rangle \right ].
\end{equation}
This is the basic formula of extracting the SZ temperature
fluctuations power spectrum of 2MASS clusters from the WMAP III
map. The result is presented in Figure 10, in which we show the
$P^{\rm top}_{\bf j}$ given by 0.11\% top modes(black points). The
scales of ${\bf j}=(8,8)$, (8,7) and (7,7) is corresponding to
angular $0^{\circ}.48$, $0^{\circ}.68$ and $0^{\circ}.96$,
respectively. The error bars are given by jack-knife. As a
comparison, Figure 10 also shows the simulation results of both
$P_{\bf j}$ (dashed lines) and $P^{\rm top}_{\bf j}$ for top
0.11\% modes(solid lines). The error bars are given by the 40
independent simulation maps using jack-knife.

The shape of all the power spectra shown in Figure 10 are similar.
Their amplitudes are, however, very different. The extracted
$P^{\rm top}_{\bf j}$ of SZ effect from the 0.11\% top modes
associated with top 2MASS DWT clusters is much higher than $P^{\rm
top}_{\bf j}$ of the 0.11\% top modes of the simulation map with
$\sigma_8=0.74$. The data points are about the same, and even
higher than $P_{\bf j}$ of sample $\sigma_8=0.74$. Since $P_{\bf
j}$ is given by all modes of the sample, while $P^{\rm top}_{\bf
j}$ is only from the top 0.11\% mode. From eqs.(3) and (17) we
must have $P_{\bf j}\geq P^{\rm top}_{\bf j}$. Therefore, the result of
sample $\sigma_8=0.74$ is unacceptable with confidence level 99\%.
On the other hand, for sample of $\sigma_8=0.84$, the extracted SZ
power spectrum (data points) is safely lower than the simulation
$P_{\bf j}$, and it is also basically consistent with 0.11\% top
modes $P^{\rm top}_{\bf j}$ of the simulation map.

Therefore, we can conclude that the power spectrum of the SZ
temperature fluctuations on sub-degree scales extracted from the
WMAP III shows a significant deviation from sample with the
parameter of density fluctuations amplitude $\sigma_8=0.74$, but
is consistent with the simulation sample produced with
$\sigma_8=0.84$, if all other cosmological parameters are the same
as that given by the three-year WMAP.

\section{Discussions and conclusions}

We studied the SZ-effect-induced temperature decrease and
fluctuations of the three-year WMAP maps caused by the 2MASS
galaxies. The DWT variables are very useful to study this problem,
as the two sets of DWT variables, given by the decompositions of
the scaling function and wavelet, are just suitable to describe,
respectively, the SZ temperature decrease and temperature
fluctuations.

We show that the maps of both SZ temperature decrease and
temperature fluctuations are highly non-Gaussian. The PDF of the
DWT local power has a long tail. More than $70\%$ power of the SZ
effect on clusters scales is given by $1\%$ top modes associated
with top clusters. On the other hand, the field of CMB temperature
fluctuations is Gaussian. One can measure the long tail events of
the SZ effect from the noisy WMAP III maps. These long tail modes
yield an useful estimation of SZ effect power spectrum. It gives,
at least, a robust lower limit of the power spectrum of the SZ
temperature fluctuations.

We found that the WMAP temperature shows significant decrease at
the positions from 100 up to $\sim$ 300 top clusters of 2MASS
galaxies. We also found that there is significant power excess at
the positions of these top clusters. This power excess is a
measurement of the SZ effect power spectrum. We find that the
power spectrum given by the 0.11\% top modes associated with top
clusters of the 2MASS galaxies, shows even higher amplitude than
that of simulation sample with $\sigma_8=0.74$. On the other hand,
the power spectrum of these top modes is bascially consistent with
$\sigma_8=0.84$ simulation SZ effect map. Therefore, the SZ effect
power spectrum on sub-degree scales seems to favor parameter
$\sigma_8>0.74$. This result supports the direct measurement of
the SZ effect power excess on arcminute scales from data of CBI
(Mason et al.2003; Readhead et al. 2004) and BIMA (Dawson et al.
2006).

\acknowledgments

This work is in partial supported by the NSF AST-0507340. We
acknowledge the use of the HEALPix software and analysis package
for producing simulation maps of the CMB. Liang Cao acknowledges
the fellowship provided by the International Center for
Relativistic Astrophysics.

\clearpage
\begin{figure}
\centerline{\includegraphics[width=12.0cm,angle=0]{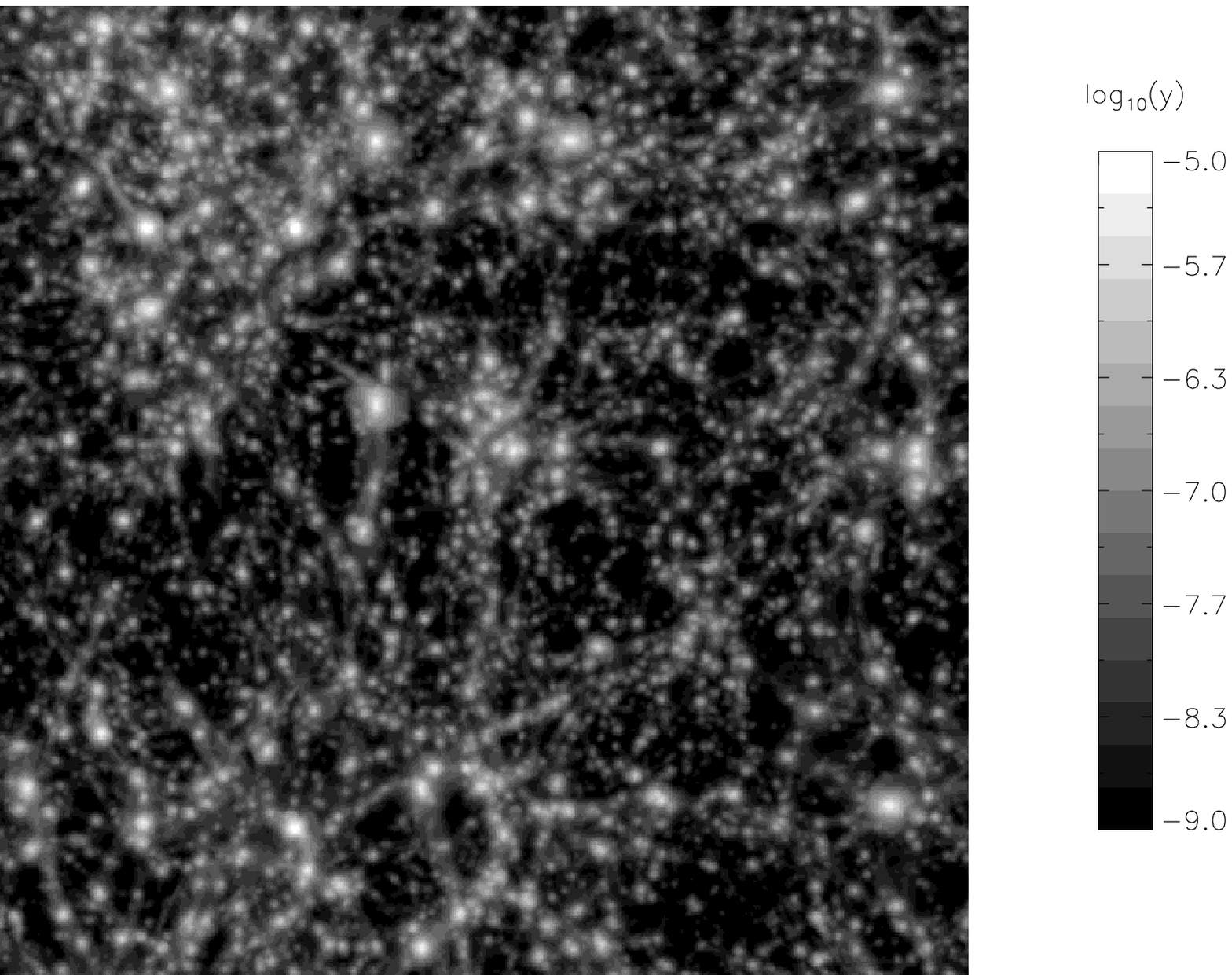}}
\vspace{2mm}
\centerline{\includegraphics[width=12.0cm,angle=0]{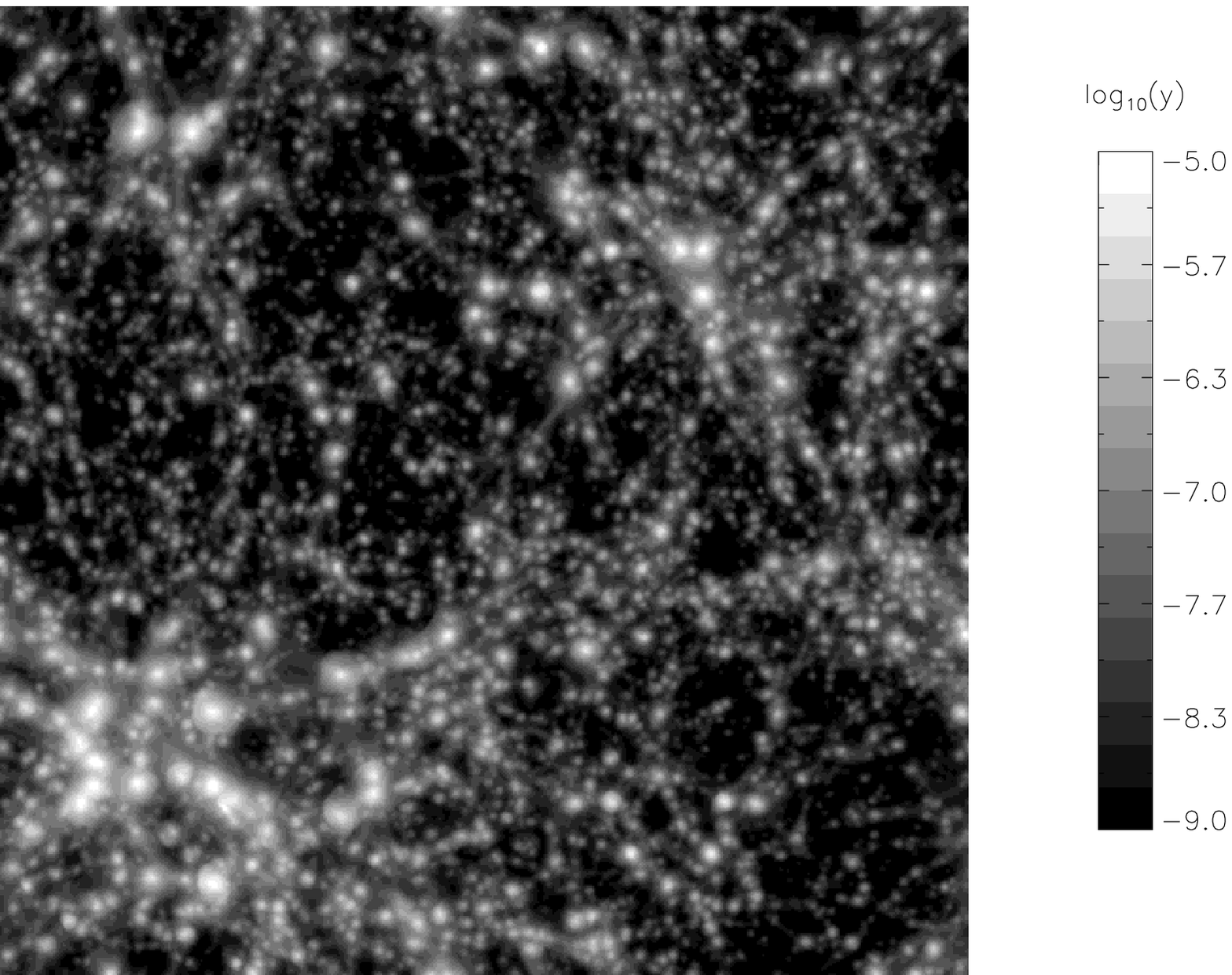}}
\caption{The SZ effect Compton parameter $y$ maps of simulation
samples with redshift $z\leq 0.1$ and $\sigma_8=0.74$(top) and
$\sigma_8=0.84$(bottom).}
\end{figure}

\begin{figure}
\figurenum{2}\epsscale{1.0}\plotone{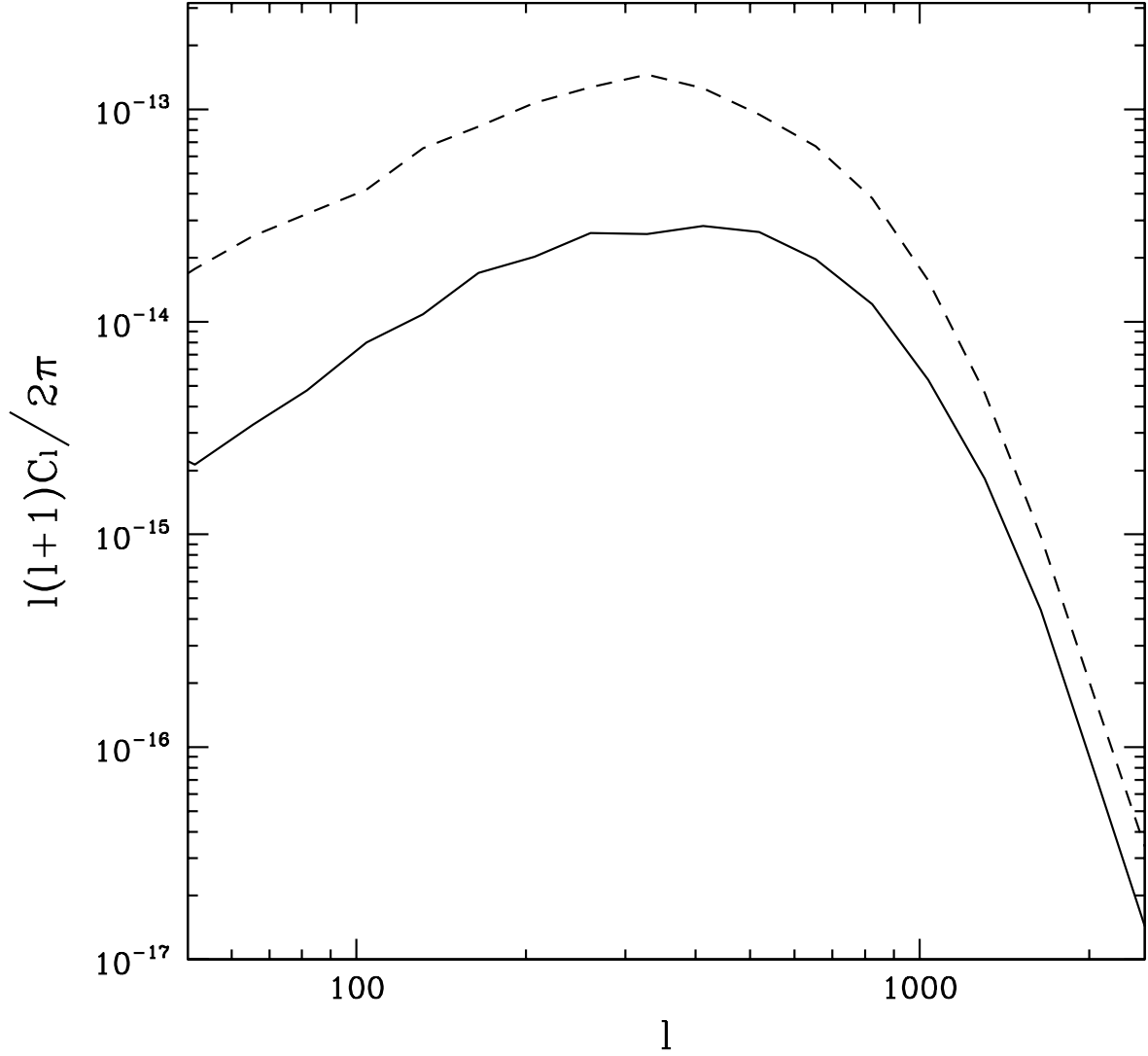} \figcaption{Angular
power spectra of SZ effect $y$ maps for samples given by
hydrodynamic simulations with parameter $\sigma_8=0.74$(solid) and
0.84(dashed).}
\end{figure}

\begin{figure}
\figurenum{3}\epsscale{1.0}\plotone{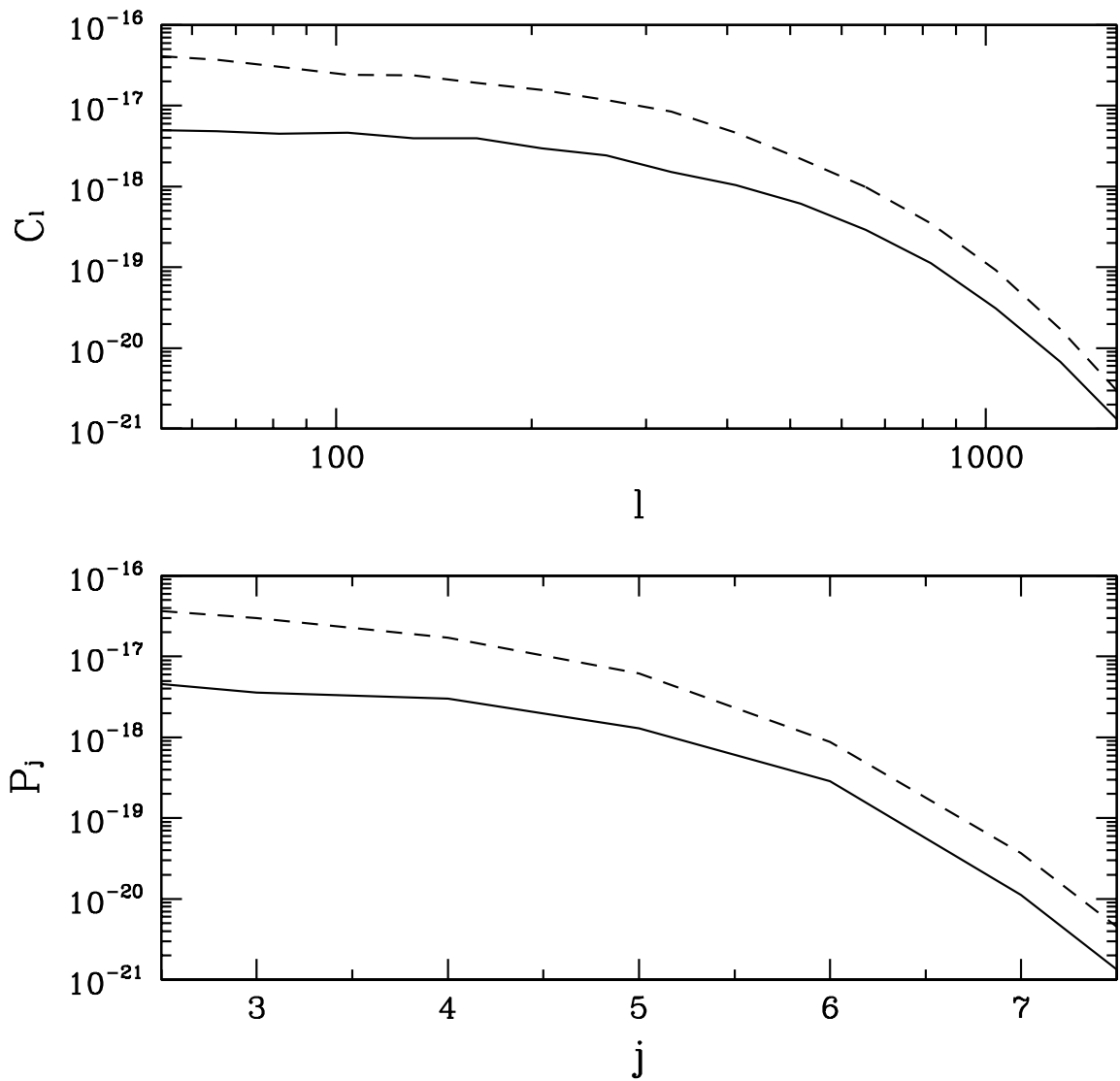} \figcaption{A comparison
between the angular power spectra of SZ effect parameter $y$
maps given by the Fourier modes (top panel) and DWT modes (bottom panel).
In each panel, the solid and dashed lines represent $\sigma_8=0.74$
and 0.84 respectively. $l$ at top panel is the index of spherical
harmonic function. Scale $j$ at the bottom panel corresponds
to angular scale of $19.1^{\circ}/2^j$.}
\end{figure}

\begin{figure}
\figurenum{4}\epsscale{1.0}\plotone{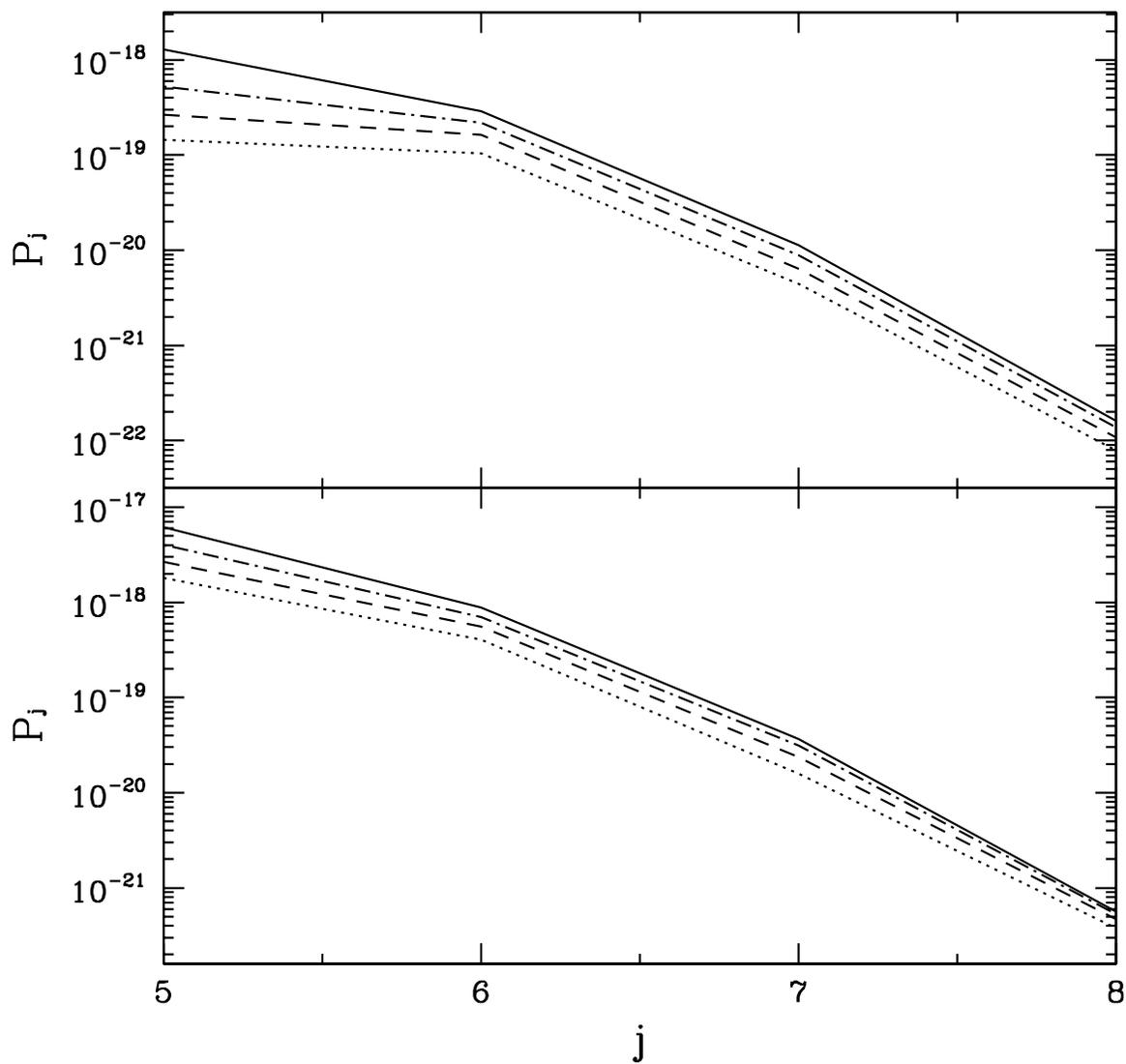} \figcaption{DWT power
spectrum of SZ effect for simulation samples of $\sigma_8=0.74$(top)
and 0.84(bottom). In each panel, we show the DWT power spectra of
whole modes (solid), and 1\%(dot-dashed), 0.3\%(dashed) and
0.1\%(dotted) modes with top power.  Scale $j$ is corresponding to
angular scale of $19.1^{\circ}/2^j$.}
\end{figure}

\begin{figure}
\figurenum{5}\epsscale{1.0}\plotone{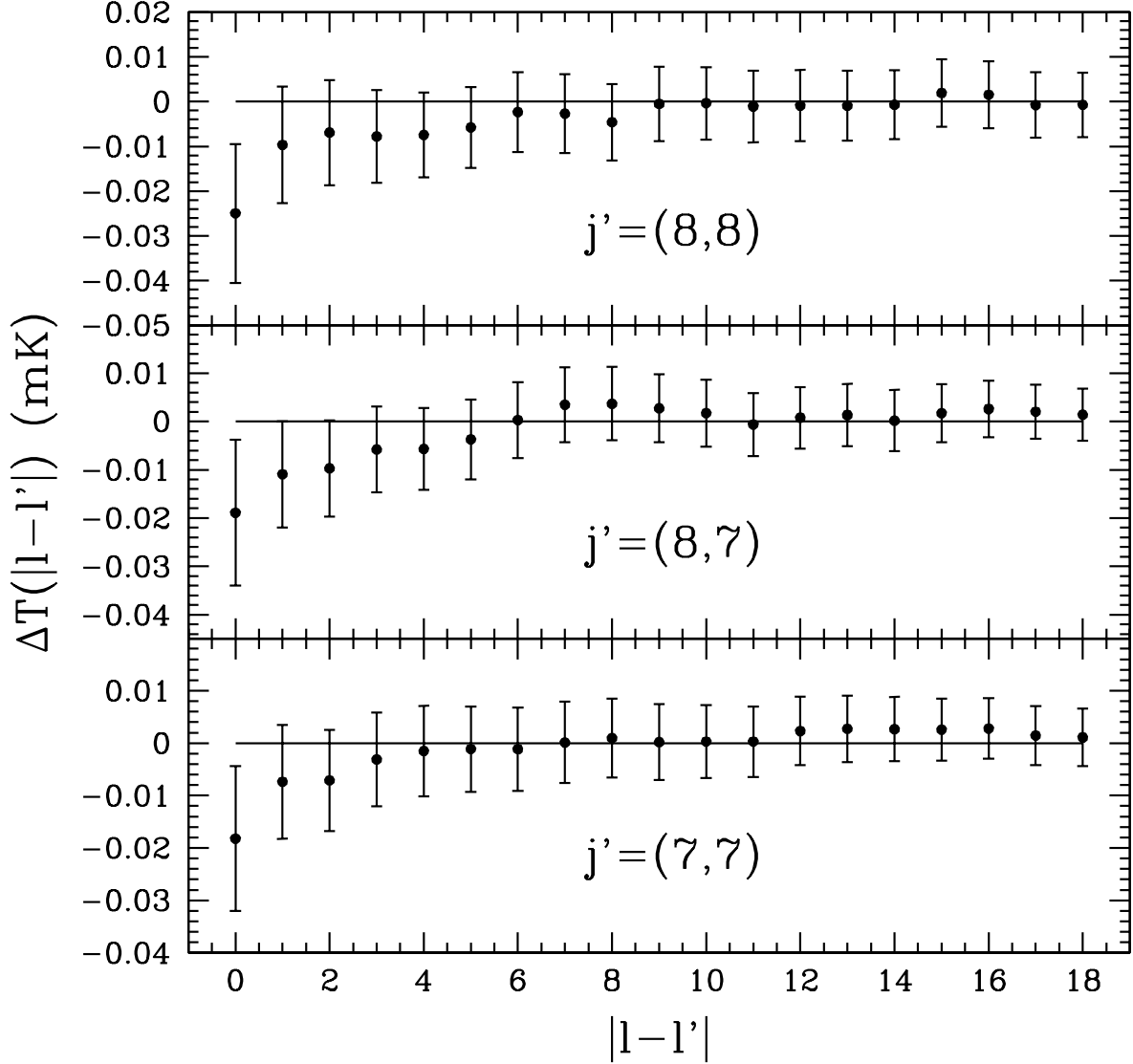} \figcaption{Cross
correlation $\Delta T^{\rm wmap-2mass}(|\bf l-l'|)$[eq.(6)] between
the top 100 ${\bf j}=(8,8)$ DWT clusters of 2MASS galaxies and
$\Delta T_{\bf j',l}$ given by the full resolution coadded 3 year
sky map for $W$ band. The sacle of $\Delta T_{\bf j',l}$ is taken to
be ${\bf j'}=(8,8)$ (top), $(8,7)$ (middle) and $(7,7)$ (bottom). In
each panel, the error bars are given by the 1-$\sigma$ error of
cross correlation between the top 2MASS DWT clusters with 100 CMB
simulations. The angular scale of $|\bf l-l'|$ is $|{\bf
l-l'}|123.88^{\circ}/2^8$ degree.}
\end{figure}

\begin{figure}
\figurenum{6}\epsscale{1.0}\plotone{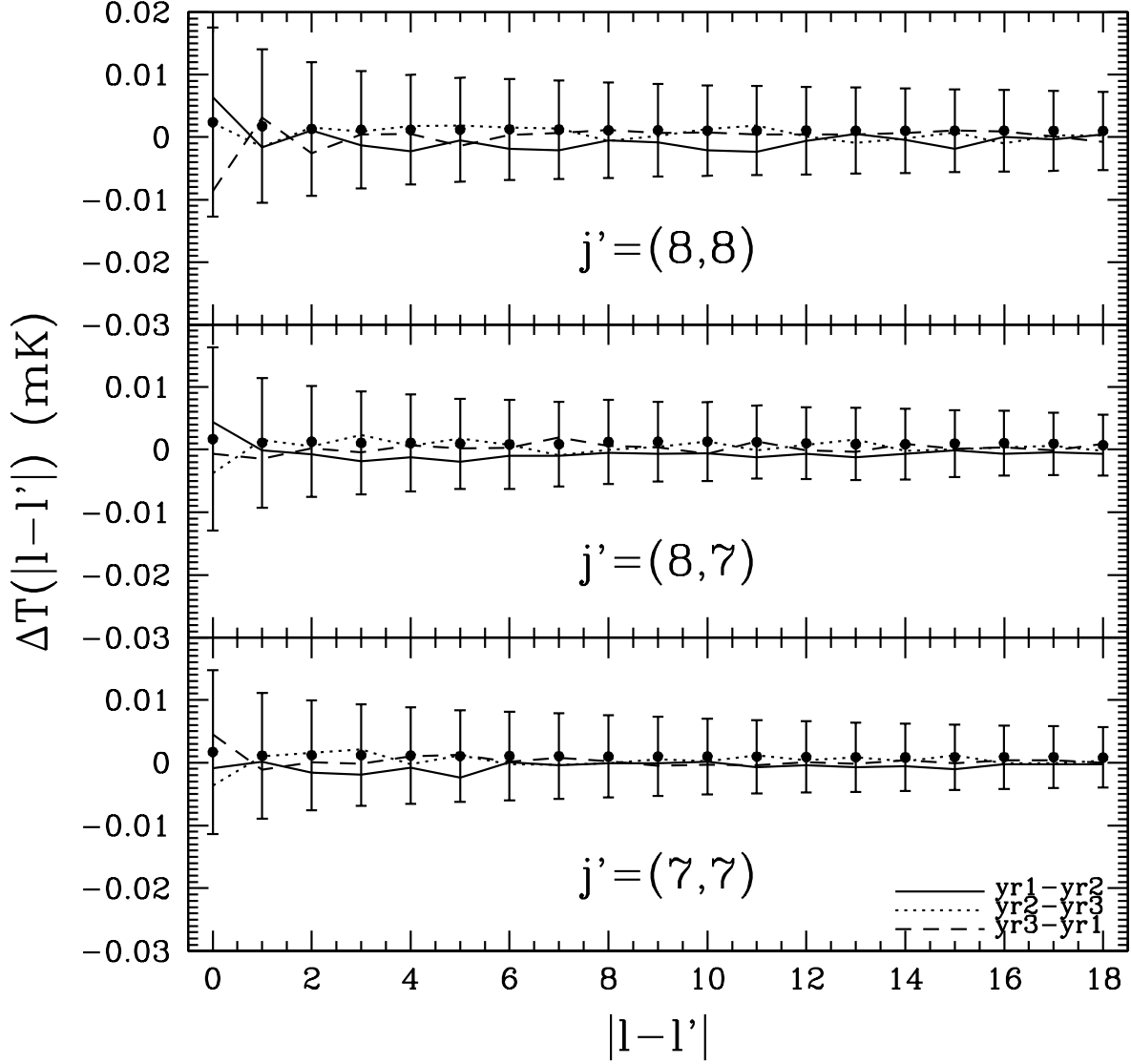} \figcaption{Cross
correlation $\Delta T^{\rm wmap-2mass}(|\bf l-l'|)$[eq.(6)] between
the top 100 ${\bf j}=(8,8)$ DWT clusters of 2MASS galaxies and
$\Delta T_{\bf j',l'}$ given by map of the differences between 1st
and 2nd years (solid), 2nd and 3rd years (dotted), 3rd and 1st years
(dashed) of WMAP data on $W$ band. The sacle of $\Delta T_{\bf
j',l}$ is taken to be ${\bf j'}=(8,8)$ (top), $(8,7)$ (middle) and
$(7,7)$ (bottom). The angular scale of $|\bf l-l'|$ is $|{\bf
l-l'}|123.88^{\circ}/2^8$.}
\end{figure}

\begin{figure}
\figurenum{7}\epsscale{1.0}\plotone{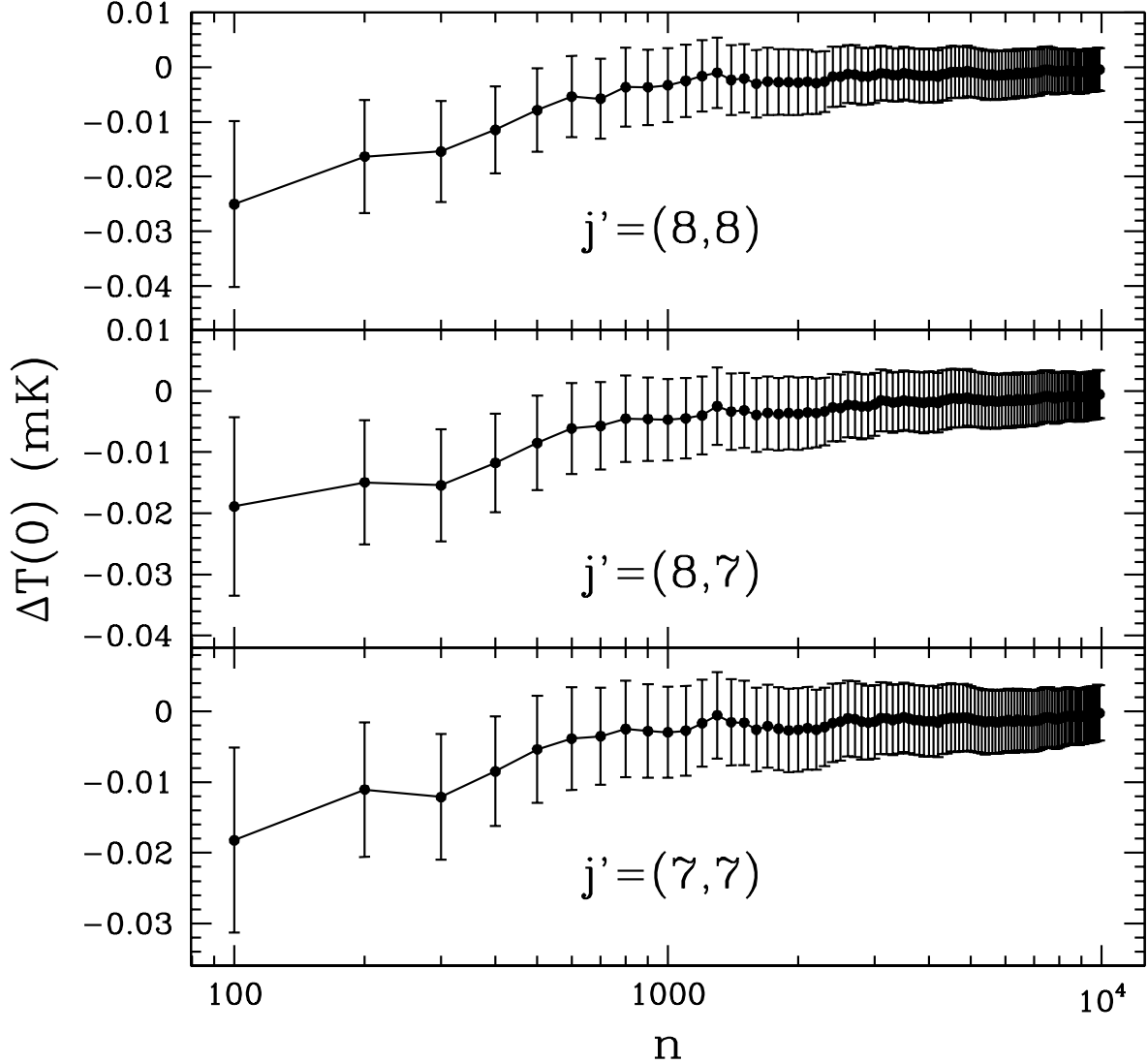} \figcaption{The cluster
richness dependence of the SZ effect temperature decrease $\Delta
T^{\rm wmap-2mass}(0)$, in which $n$ is the number of top ${\bf
j}=(8,8)$ clusters of 2MASS galaxies. The scale of $\Delta T_{\bf
j',l}$ is taken to be ${\bf j'}=(8,8)$ (top), $(8,7)$ (middle) and
$(7,7)$ (bottom). The error bars are given by the 1-$\sigma$ error
of cross correlation between the top 2MASS DWT clusters with 100 CMB
simulations.}
\end{figure}

\begin{figure}
\figurenum{8}\epsscale{1.0}\plotone{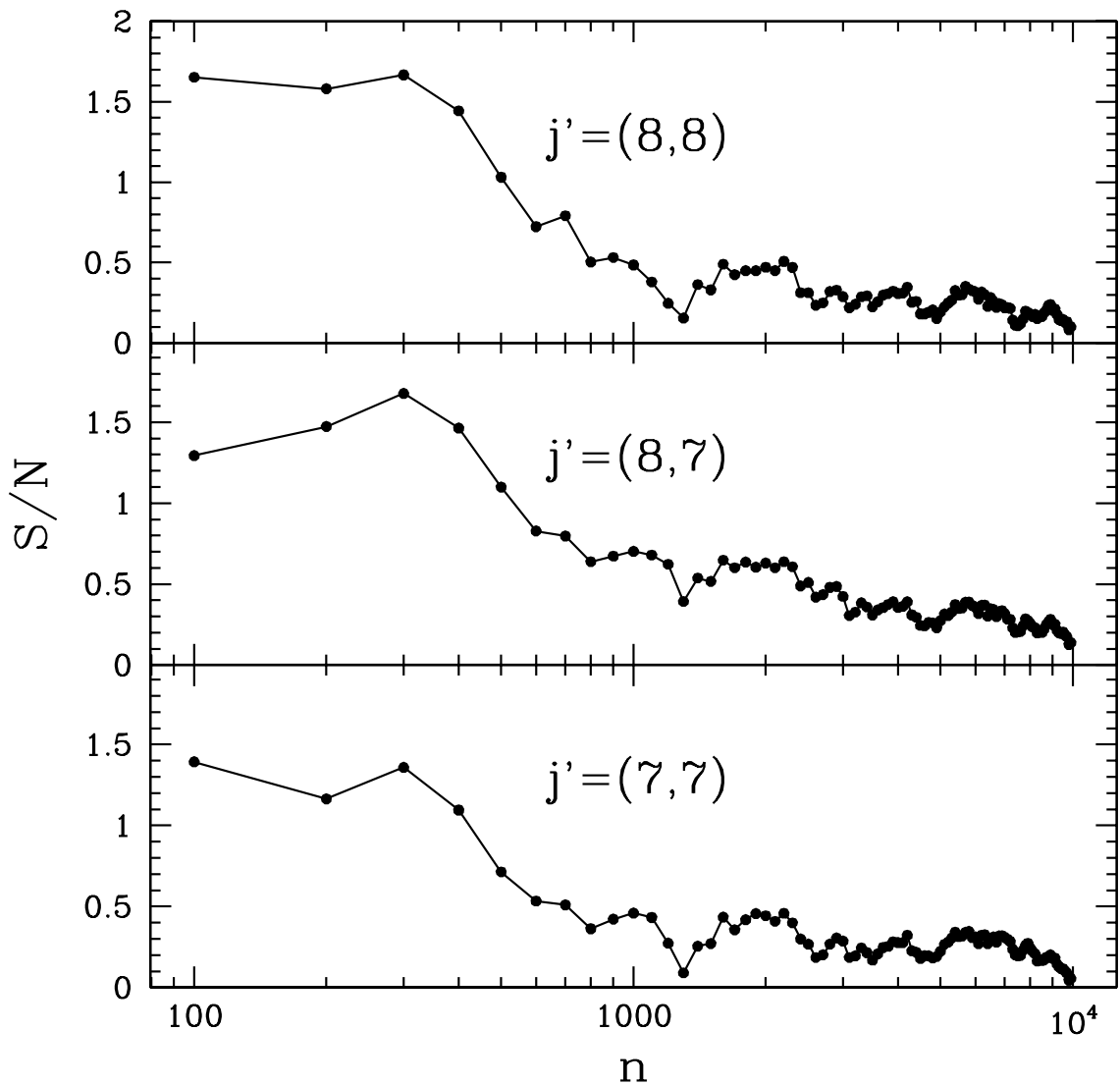} \figcaption{The
signal-to-noise ratio of the SZ effect temperature decrease
$\Delta T^{\rm wmap-2mass}(0)$ against clusters richness, in which
$n$ is the number of top ${\bf j}=(8,8)$ clusters of 2MASS
galaxies. The scale of $\Delta T_{\bf j',l}$ is taken to be ${\bf
j'}=(8,8)$ (top), $(8,7)$ (middle) and $(7,7)$ (bottom).}
\end{figure}

\begin{figure}
\figurenum{9}\epsscale{1.0}\plotone{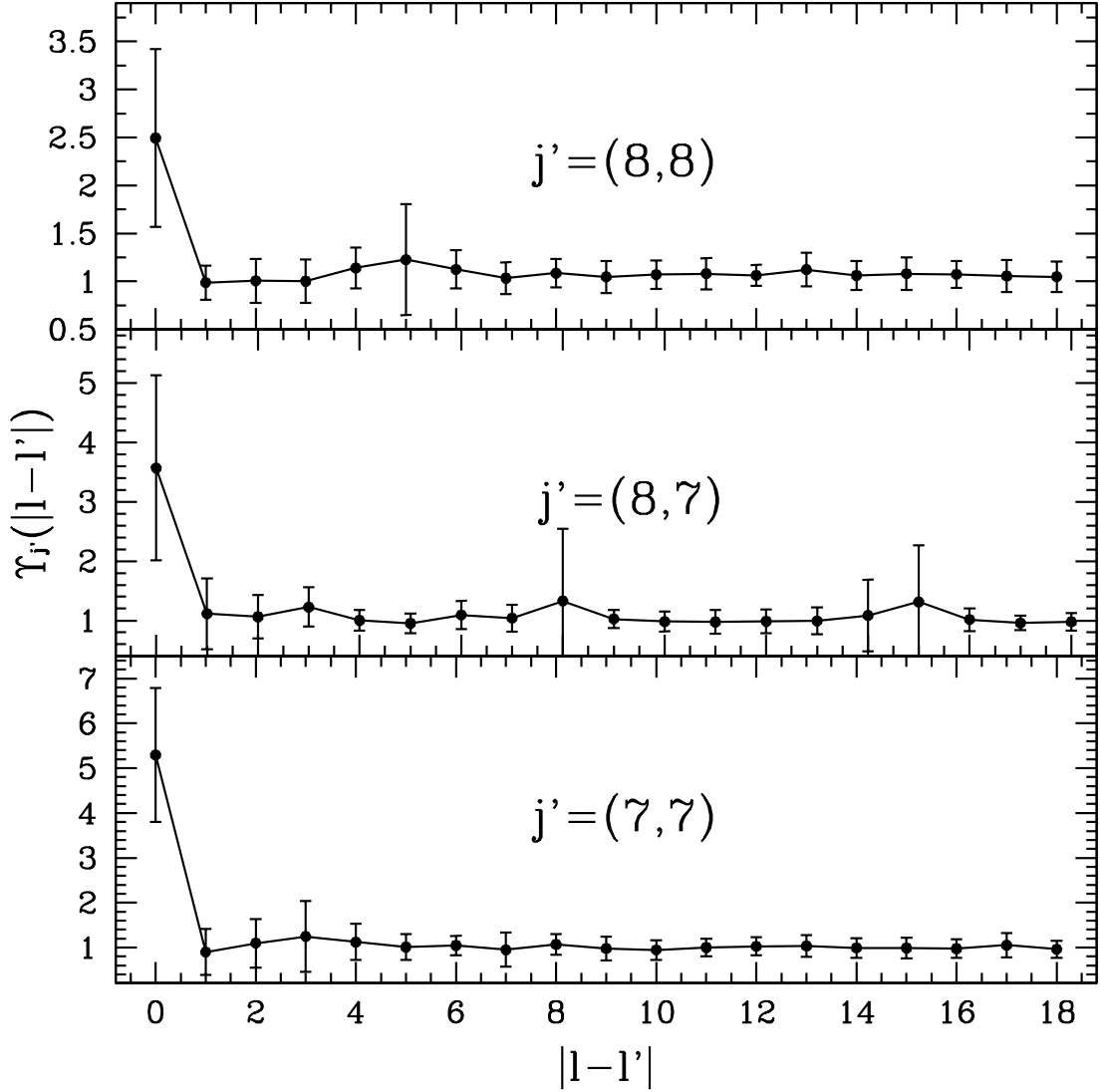}
\figcaption{$\Upsilon_{\bf j'}(|\bf l-l'|)$ [eq.(16)]. The average
is taken over 0.11\% modes with top $|\tilde{\epsilon}_{\bf
j',l'}|^2$ in the positions of 300 top cluster with ${\bf j}=(8,8)$.
The scales ${\bf j'}$ is taken to be  ${\bf j'}=(8,8)$ (top), ${\bf
j'}=(8,7)$ (middle) and ${\bf j'}=(7,7)$(bottom). The angular scale
of $|\bf l-l'|$ is $|{\bf l-l'}|123.88^{\circ}/2^8$.}
\end{figure}

\begin{figure}
\figurenum{10}\epsscale{1.0}\plotone{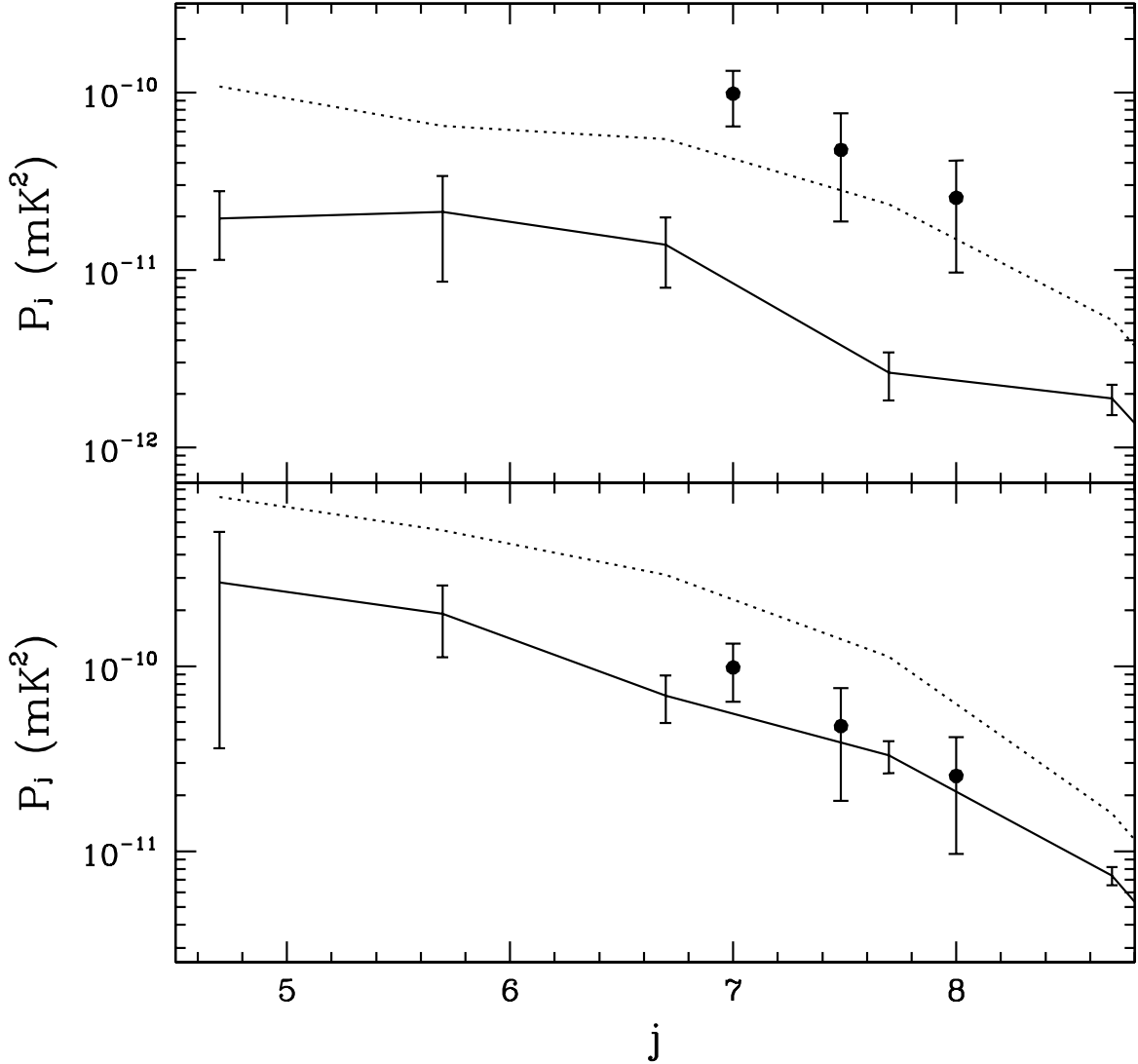} \figcaption{The DWT
power spectrum of SZ effect temperature fluctuations at W band.
The results from eq.(18) with 0.11\% top modes associated with the
top DWT clusters is shown by black point with error bar. The
dashed and solid lines show the power spectrum of simulation
sample with all modes and 0.11\% modes respectively. The top panel
is for sample of $\sigma_8=0.74$, the bottom is for
$\sigma_8=0.84$. Scale $j$ is corresponding to angular scale of
$123.88^{\circ}/2^j$. The three black points of scales of ${\bf
j}=(8,8)$, (8,7) and (7,7) have angular scale of $0^{\circ}.48$,
$0^{\circ}.68$ and $0^{\circ}.96$, respectively.}
\end{figure}
\end{document}